\documentstyle[floats,prb,epsf,aps]{revtex}
\newcommand{\ba}{\begin{eqnarray}}
\newcommand{\ea}{\end{eqnarray}}
\newcommand{\bea}{\begin{eqnarray}}
\newcommand{\eea}{\end{eqnarray}}
\newcommand{\be}{\begin{equation}}
\newcommand{\ee}{\end{equation}}

\newcommand{\nn}{\nonumber}

\begin{document}
\flushbottom
% ===================================================================
\wideabs{
\draft
\title{Density of states, screening and the width of the quantum Hall
      plateaus}  
\author{K. Tsemekhman, V. Tsemekhman}
\address{Department of Physics, University of Washington,
 Box 351560, Seattle, Washington 98195}
\author{C. Wexler}
\address{Department of Physics, Box 118440, University of Florida,
 Gainesville, Florida 32611}
\date{7 Oct 1998}

\maketitle
% ===================================================================
\begin{abstract}

We present a consistent treatment of the quantum Hall effect within 
the electrostatic approximation.  
We derive the form of the density of states (DOS) which differs from
the usual gaussian shape valid for non-interacting electrons. 
Below a crossover temperature, this DOS gradually
transforms into the non-interacting DOS near the Fermi energy.
We obtain an estimate of this crossover temperature, which
determines the range of validity of this framework, and allows
a reconciliation of the width and temperature dependence of the
quantum Hall plateaus with experiments.
The DOS dramatically enhances the electron-lattice relaxation rate,
thus reducing the possibility of electron overheating.  

\end{abstract}
} % end of wideabs
% ===================================================================

% ===================================================================
\section{Introduction}

The quantum Hall effect (QHE) is usually thought of as a
manifestation of the absence of scattering of electrons carrying the
Hall current. This is caused by the Fermi level falling between two
mobility edges. Each mobility edge in an infinite disordered system is
the energy of a unique extended state on a given Landau level (LL). 
It has been shown, numerically and in a number of approximations
\cite{Ando1,Wegner1,Huckenstein1}, that the density of states (DOS) of
non-interacting electrons is symmetric around the extended state
energy of each LL, and has a width that depends on the strength of
disorder but which is always much smaller than the cyclotron energy
${\hbar \omega_{c}}$,  with ${\omega_{c}=eB/mc}$. This density of
states proved to be consistent with the qualitative explanation of the
widths of QH plateaus \cite{Prange1} and with the numerical values of
critical exponents describing the scaling at phase transitions between
QH plateaus \cite{Huckenstein1}. Interaction between electrons has,
however, been left out of those treatments, without a strong 
{\it a priori} justification.

On the other hand, Efros \cite{Efros} has suggested that, in presence 
of long-range fluctuations of the disorder potential, interactions
between electrons are crucial, leading to collective screening of the
disorder potential. Efros showed that, if the electrostatic
approximation is appropriate (which is true when the screening radius
is much smaller than any other relevant length), there could be 
two types of screening which he denoted ``linear'' and ``non-linear''.
Linear screening is essentially metallic-like: in a region with linear
screening the energies of two-dimensional electrons are confined to an
interval of the order of temperature around the chemical potential and
states on the top LL are partially filled.  
On the contrary, non-linear screening is much less effective. In such
regions the amplitude of fluctuations of the disorder potential is
reduced due to electric charge of electrons, but this reduction is
limited by the maximum density of electrons on the LL and by the inter
LL energy gap. 

The existence of linear screening is supported by the theory of
compressible edge strips in a gate-confined two-dimensional electron
system (2DES) \cite{Chklovskii}. Although the nature of the potential
being screened in these long compressible strips is quite different from
that of the disorder potential, the difference in the description of
screening in these two cases is relatively small. Conditions for the
validity of the electrostatic approximation itself, on the other hand,
should be universal. If such conditions are satisfied, one can expect
a considerable fraction of the area of the 2DES to be covered by 
compressible regions and, therefore, a large density of states near the
Fermi level (and at energies differing by $\hbar \omega_c$). Such DOS
(we  will refer to it as the {\em screened} DOS) is completely
different from the corresponding DOS for non-interacting
electrons. The difference leads to appreciable differences in the
predicted widths of QH plateaus \cite{Efros}, apparently, for the
rates of various scattering processes. Similar considerations were also
successfully applied in the past to a quantitatively correct theory of
the current breakdown of the QHE \cite{We_BKDN}.

Dissipative transport properties at sufficiently high temperatures are
due to thermal activation of electrons to unfilled LL (or due to
holes in filled LL). It was shown both experimentally
\cite{klitz1,Komiyama1} and in a number of theoretical models
\cite{Shklov1,Shklov2,Kivelson1} that the average dissipative
conductivity of the 2DES in the quantum Hall regime satisfies the
equation
\be
    \sigma_{xx}=\sigma_{0} \, {\exp(-\epsilon_{a}/kT_{e})}\;,
\label{activ}
\ee
where $\sigma_{0}$ is generally a non-universal constant ranging
between $0.5 e^2/h$ and $5 e^2/h$, $\epsilon_{a}$ is the activation
energy defined as the distance between the Fermi level and the energy
of the {\it extended} state, and $T_{e}$ is the temperature of
electron gas. We keep the subscript in $T_{e}$ to distinguish it 
from the lattice temperature $T_{b}$ in cases when these are
different. The range of validity of Eq.\ (\ref{activ}) will be
discussed later in this paper. We want to stress at this point,
however, that Eq.\ (\ref{activ}) implies an {\it exponential}
smallness of $\sigma_{xx}$, and therefore, exponentially small Joule
heating due to scattering inside the system. 

One of the consequences of the non-interacting DOS is the suppression of
scattering events by delocalized states since they are all deep
under the Fermi level. Therefore, the energy relaxation rate via
phonon emission would also be small in this case. If this non-interacting
approximation was valid, under some conditions the phonon-mediated
energy relaxation could become slower that the Joule heating,
resulting in observable overheating of the 2DES
\cite{klitz1,Komiyama1,Mintz,Nachtwei}. Alternatively, the interacting
DOS provides for a macroscopic fraction of states at the Fermi level
and, therefore, allows a very effective coupling of the 2DES to the
lattice. Overheating, in this case, becomes negligible.

In this paper we address the question of the range of validity of
both non-interacting DOS and screened DOS. The former is
valid at low temperatures, while the latter dominates in the higher
temperature regimes. We present an estimate of the crossover
temperature falling in the range between 10 mK and 1K depending on
various parameters of the sample and on experimental quantities.
At temperatures above this crossover value, the screened DOS leads to
a gigantic enhancement of the typical phonon emission rates and,
therefore, to the absence of overheating in the QH regime.

The paper is organized as follows. 
In Sec.\ \ref{sec:screening} we discuss the screening of a
long-range disorder potential, closely following the formulation by
Efros \cite{Efros}. The concept of screening is then used to introduce
the new form of the DOS of interacting electrons. In Sec.\ \ref{range}
we examine the range of validity of the electrostatic approximation
and present an intuitive picture of the physics beyond this
range. Section \ref{fintemp} deals with the evolution of the quantum
Hall effect with temperature. We derive an estimate for the
crossover temperature below which the DOS behaves more closely to the
conventional DOS for non-interacting electrons in a random potential,
at least close to the Fermi energy. We also propose a theory of the
melting of the QH liquid and present numerical results supporting our
understanding of the development of the DOS with the
temperature. Finally, in Sec.\ \ref{heat} we discuss dissipative 
processes in the context of the screened DOS, which dramatically
enhances the electron-lattice relaxation rate thus eliminating 
electron overheating in the QH effect.

%********************************************************************
\section{Screening}
\label{sec:screening}

%In this Section we show that the large width of the QH plateau
%is produced by the short range disorder potential. The long range disorder
%results in much narrower plateaus but has very distinct features
%from the short range disorder density of states in 2DES. Which of the
%two types of disorder prevails in a given sample depends entirely on
%the ratio of temperature and amplitude of the short-range disorder
%potential.

We will describe the QH system in a hierarchy of approximations. In
each hierarchical level the common goal will be the
determination of the spatial distribution of filling factors for each
LL [$n^{(N)}({\bf r})$]. In a strong magnetic, for sufficiently slowly
varying external potentials (of any nature) it is a good approximation
to consider all the states on consecutive LLs to be locally
separated by $\hbar \omega_{c}$. 

The hierarchy starts with a simple electrostatic approximation. A
local filling factor is used to obtain the classical charge
density distribution $\rho({\bf r})$ using
\be
        \rho({\bf r}) = \frac{1}{2\pi l^{2}} 
        \sum_{N}{n^{(N)}({\bf r})} \;,
\ee
where $l \!=\! (\hbar c/eB)^{1/2} \!\sim\! 100$ {\AA } is the magnetic
length. This approximation is equivalent to finding the electron density
distribution $\rho({\bf r})$ which minimizes the free energy of the
system. At zero temperature this corresponds to the minimization of a
Hamiltonian consisting of the potential energy of interactions between
electrons, interaction with the background disorder and quenched
kinetic energy:
\ba
        {\cal H}_1 =&& \frac{e^2}{2 \epsilon} \int \! \! \! \int \!
         d^{2} \!{\bf  r} \, d^{2}
          \!{\bf r'} \: \frac {\rho({\bf r}) \rho({\bf
              r'})}{ \mid {\bf r} - {\bf r'} \mid} + 
          e \int \! d^{2} \!{\bf r} \, \rho ({\bf r}) V( {\bf r}) \nn \\
&&+          \sum _{N} (N + \frac{1}{2}) \hbar \omega_{c}  \int \!
          d^{2} \! {\bf r} \, n^{(N)} ({\bf r}) \;.
\label{Elst}
\ea
The third term in the right-hand side of Eq.\ (\ref{Elst}) reflects
the existence of a gap between the Landau levels. The problem 
corresponding to Hamiltonian ${\cal H}_{1}$  was first analyzed and
then solved numerically by Efros and co-workers
\cite{Efros,Efros3,Efros4} for a disorder potential created 
by remote donors in a plane parallel to the 2DES.
As these authors pointed out, for essentially any disorder potential
the response of the electron system results in two types of screening
which they labeled {\em linear} and {\em non-linear}. In the linear
screening regions the total electrostatic potential 
\be
        \Phi ({\bf r}) = 
        \frac{e}{\epsilon} \int \! d^{2} \!{\bf
          r'}
          \: \frac {\rho({\bf r'})}{ \mid {\bf r} - {\bf r'} \mid} + 
          e V( {\bf r})
\label{Potential}
\ee
is flat and satisfies the equation
\be
        \Phi({\bf r}) = \mu - 
         (N + \frac{1}{2}) \hbar \omega_{c}\;,
\label{Compressible}
\ee
where $\mu$ is the chemical potential and $N$ is the top 
partially-filled LL in the vicinity of point ${\, \bf r}$. The filling  
factor $n^{(N)}({\bf r}) $ in such areas, which are called 
{\it compressible}, varies between zero and one. 

In contrast to this situation, there are also {\it incompressible}
regions of non-linear screening which are characterized by a complete 
filling of the top-most Landau level $n^{(N)}({\bf r}) \! =\! 1$. In
these cases the potential satisfies the inequalities   
\be
        \mu - (N + \frac{3}{2}) \hbar \omega_{c} \; < \; 
        \Phi ({\bf r}) \; < \;
        \mu - (N + \frac{1}{2}) \hbar \omega_{c} \;.
\label{Incompressible}
\ee

Before going up in hierarchy, we give our interpretation of these
equations. With respect to the effectiveness of screening, one can
imagine two situations: a compressible region percolates or
an incompressible region percolates. In the former situation, there
are extended states at the Fermi energy and the system behaves
essentially as a metal. The latter scenario describes the
quantum Hall effect since in this case, the extended states are away
from the Fermi level, scattering processes are suppressed and the Hall
conductance is quantized with a very high accuracy. 

\begin{figure}
        \begin{center}
        \leavevmode
        \epsfbox {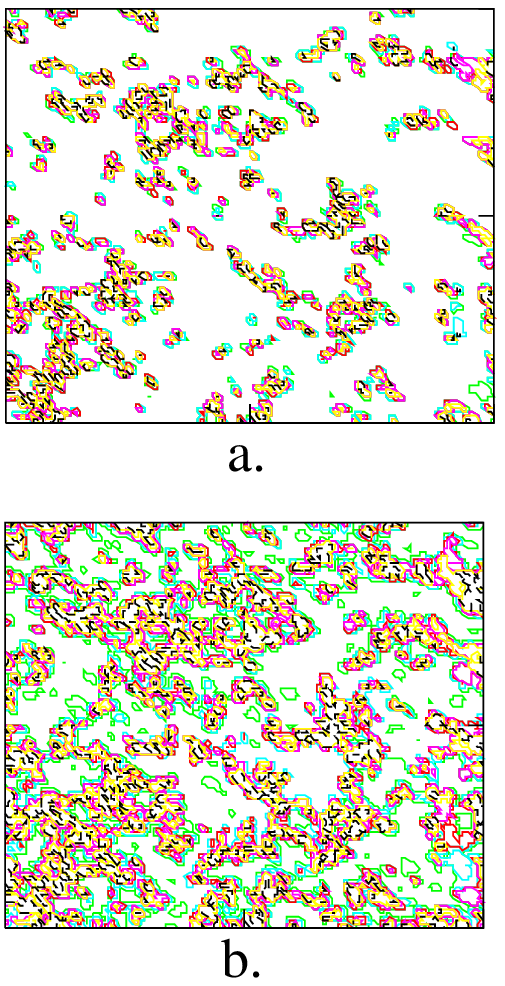}
        \end{center}
 \caption[Electron density distribution in the 2DES] 
{Electron density distribution in the 2DES; shaded areas
  represent compressible, partially filled states on the first or
  second LL, white regions are incompressible. a.: Incompressible
  region percolates (QH regime); b.: Compressible region percolates. 
  {\label{Example}}}
\end{figure}

A typical picture of an inhomogeneously filled 2DES is shown on 
Fig. \ref{Example}. It is the result of our numerical simulations of
the 2DES at finite temperature, the essence of which will be described
in the following Sections. In the QH regime (Fig.\ \ref{Example}(a))
the majority of the plane is taken by incompressible (completely
filled LLs) areas. Yet, compressible lakes still take a finite
fraction of the sample.
By the virtue of Eq.\ (\ref{Compressible}) we conclude that a finite
fraction of the total number of states have energy equal to the chemical
potential. Since all the partially filled regions have sizes
smaller than the sample size, the corresponding quantum states ought
to be localized and do not contribute to dissipation. To a limited
extent, the situation is similar to a common 3-dimensional insulator
system, 
where there are states present at the Fermi level but all of them are
localized in space. However, there is a major departure from this
picture: while in a common insulator the density of states is
smooth around the Fermi level (or even vanishes in semiconductors due
to the Coulomb gap), in the case under study it is strongly peaked around
the Fermi level. Finite temperature and neglected correlations, the
effects which we consider below, tend to smear the peak, but the 
weight of partially filled states remains finite.

%\end{multicols}

%\begin{multicols}{2}

\begin{figure}
  \begin{center}
    \leavevmode
    \epsfbox {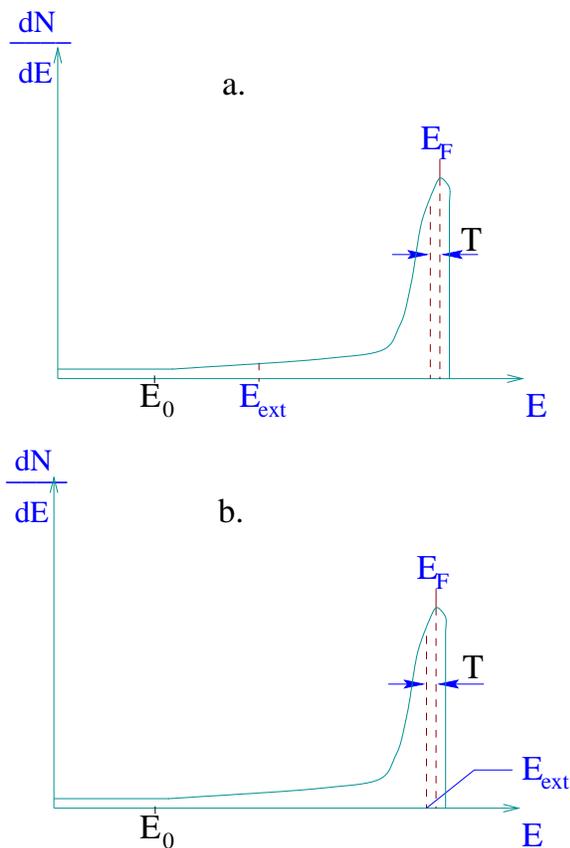}
%       \{epsffile{}
    \caption[Density of states for interacting electrons] 
    {Density of states for interacting electrons at two different
values of the magnetic field. a.:\, In the quantum Hall regime; b.:\,
In transitional regime. $E_0$ is the center of LL in the corresponding
non-interacting system. Note that the only visible difference is the
position of the energy of the extended states relative to the Fermi
level. The DOS is always peaked at the Fermi level.
      {\label{DOS}}}
  \end{center}
\end{figure}

Quite remarkably, the form of the DOS does not change dramatically with the
onset of the dissipative regime with a percolating compressible
region [see \ref{Example}(a),(b) and Figs.\ \ref{DOS}(a),(b)]. While
the landscapes of filling factor and potential do undergo major
changes with the density of electrons passing through 
the critical value corresponding to the boundary between the two
percolating regimes, the DOS changes only quantitatively and in a smooth
fashion. The existence, in both regimes, of a finite fraction of states
on the Fermi level, allows a percolation transition to occur without a
visible transformation of the DOS. During the transition, most of the
states in the  {\it percolating} compressible region remain
localized. The presence of these localized states 
{\em at the Fermi level}, however, does have profound effects since
these states can actively participate in low energy scattering
processes, giving rise to finite dissipative conductivity. 

Comparing the situation above to the metal-insulator transition in
3-dimensional systems, we note that the latter one occurs when the
states on the Fermi level transform from localized to extended: the
number of such states in the localized regime is so small that they
are always spatially separated from each other by wide potential
barriers and can only conduct via exponentially small hopping.  
One can say that in conventional 3D metal-insulator transition the
Fermi level moves with respect to the DOS; in the 2D system under
consideration  both the Fermi energy and the DOS are almost
rigid. Instead, the main transformation is the motion of the energy of
the extended state, which approaches the Fermi energy as the system
moves {\em away} from the QH regime.

We conclude that the structure of the DOS near the Fermi level, in a
regime when the electrostatic approximation is justified, plays a
crucial role in the mechanisms of conductivity. The origin of this DOS
can easily be seen from the Hamiltonian of Eq.\ \ref{Elst}. Compared
to a typical 3-dimensional electron system, its kinetic energy term is
quenched by the strong magnetic field and replaced by the quantized
energy of electrons on a given LL. As long as this replacement is
valid and the number of filled Landau levels is sufficiently small,
there are no further degrees of freedom for the electron cloud other
than the potential energy. In 3D systems the states with the same
average potential energy (in the same spatial region) would have
different average kinetic energies thus providing for a continuous
spectrum of the energies of these quantum states; 
{\it no} preference in the DOS would then be given to the Fermi
energy. In the 2DES in a strong magnetic field, the potential energy
is minimized and all the states in the ``screened'' region tend to have
the same {\it total energy}, equal to the Fermi energy.

% ===========================================================
\section{Range of validity of the electrostatic approximation}
\label{range}

Let us analyze more carefully the assumptions of the electrostatic 
approximation. We begin by estimating the size of the neglected
kinetic energy term. Consider a mean-field equation for a single
electron in the presence of the random potential $V({\bf r})$ . Under
the assumption of a sufficiently weak and slow varying electric field
${\cal E} \! \ll \! \hbar \omega_{c} / e l$ one can expand locally
$V({\bf r})$ and write the corresponding one-particle Hamiltonian
keeping only the linear term:  
\be
   H = \frac{({\bf p} + e {\bf A}/c)^2}{ 2m} - 
   e \, {\cal E} \,y \;,
\label{Quadratic}
\ee
where ${\cal E} \!\sim \!u_0/e\lambda$ is a typical local electric field
due to a disorder potential of size $u_0$ and correlation length
$\lambda$. It is natural to choose the direction perpendicular to the
local electric field as a quasi-translationally invariant one. 
Then, within the region where this approximation is valid, it is
possible to enumerate the solutions by the wave-vectors corresponding
to the quasi-translationally invariant direction, as it is done in a
common Landau equation \cite{L&L_QM}. 

This electric field produces a quadratic energy shift which can be
referred to as the kinetic energy, since it is quadratic in the drift
velocity ${\cal E}/B$. A simple estimate gives the value 
$(e {\cal E} l)^2 / \hbar \omega_c \! \sim \! 0.01 \, \hbar \omega_c$
for long wavelength disorder potential.
This justifies neglecting the kinetic energy term in the system we are
interested in, since fluctuations of the bare potential commonly
exceed $\hbar \omega_{c}$. Moreover, in screened regions, the
minimization of the  potential energy leads to a simultaneous
reduction of the kinetic term as well (decreasing of the potential
fluctuation of a certain wavelength implies a decrease of the
corresponding electric field by the same factor). Therefore, inclusion
of the quadratic kinetic energy term does not produce any major
change, which validates the electrostatic approximation used in Sec.\
\ref{sec:screening}.

The electrostatic picture can still be changed by quantum
effects. Quantum correlations contribute an additional term into the
energy density of the homogeneous 2DES \cite{Efros3,Fano}. One
possible form of writing this term is \cite{Efros3}: 
\be
        H_{0}(\nu) = \sqrt{2\pi} \; \frac {e^2}{\epsilon} \,
        n_{0}^{3/2} \,g(\nu) \;,
\label{Correl}
\ee
where $n_{0}$ is the density on each Landau level, $\nu$ is the
filling factor and $g(\nu)$ is some complicated function of $\nu$ of 
order of unity at $\nu \! = \! 1$. This term is responsible for a
finite screening radius $r_{s}$, contrary to the classical value
$r_{s}\!=\!0$. It is clear that there are only two characteristic
length scales where quantum effects set in: the size of 
the wavefunction and the average distance between particles on each
Landau level. The first length is entirely determined by the magnitude
of magnetic field while the second one depends also on the density. 
When the filling factor is close to $\nu\!=\!1$ both length
scales are of the order of magnetic length; the screening length
should, therefore, be of the order of the magnetic length as
well. This agrees with an estimate for $q_{s}\!=\!2\pi/r_{s}$ from
Eq.\ (\ref{Correl}): 
\be
        q_{s} = 2 \pi \frac {e^2}{\epsilon}\left( \frac {d^{2}
            H_{N}}{d n^{2}} \right)^{-1} \;.
\label{Qs}
\ee

Although Eq.\ (\ref{Correl}) may not be a good approximation for such
an inhomogeneous system as the one we study, it gives a good
order-of-magnitude estimate, at least for external potentials
with wavelengths $\lambda \! > \! l$. The meaning of $q_{s}$ here is 
the same as in the usual 3-dimensional theory of screening: it measures
how effectively a given harmonic $V_{q}$ of an external potential is
reduced due to electron interactions. In the 2DES with Coulomb
interaction in all the Fourier components of external potential one
has to make a replacement $q \rightarrow q + q_{s}$ in order to
reproduce the corresponding harmonic of the screened potential. The
Fourier component of the screened potential is $\tilde{F}_{\bf q}
\!\sim \!(q + q_{s})^{-1}$,
and therefore screening is effective for harmonics with  
$q \! \ll \! q_{s}$ and  much less so for $q \! \geq  \! q_{s}$.   

The effect of electrostatic screening is the result of a {\it density}
response while the correlation energy is determined by the 
{\it filling factor} of quantum states. When the density and the
filling factor are equal to each other, that is, when the relevant
length scales are much larger than the extent of the wavefunction,
electrostatic and correlation energies behave coherently. In 
regions with {\it linear screening}, the one-particle quantum states 
that correspond to the minimum total energy of the system, have 
energies equal to the chemical potential. Although electrostatic
screening is limited by the effects of correlations, and therefore,
the electrostatic potential differs significantly from the chemical
potential, individual energies, being the
combinations of electrostatic and correlation contributions, fall
around the chemical potential ($\pm k_B T$). Such states are
filled partially and behave like a metal. 

We would like to point out that even in the regions of partial filling
there is still a residual electric field due to the finite screening
length. For disorder potentials with $q \!>\! q_{s}$ the {\it electrostatic
interaction} becomes ineffective. The range of wavevectors $q$ where,
in turn, the {\it correlation potential} becomes unable to alter the
individual particle energies prescribed by electrostatic potential
is defined by the condition that there be one particle
within one fluctuation of the potential. We come to the conclusion
that both electrostatic interaction and correlation potential do not
affect individual particle energies for external harmonic potentials
with the wavelengths $\lambda \!\le\! l$. Rephrasing it, if the system
``knows'' about the presence of fluctuations of potential with $\lambda
\! \le \! l$, the states in this system are either completely filled or
empty; no partial filling is possible. This sets a natural limit to
electrostatic approximation: if the {\it external} potential
fluctuations with the wavelength $\lambda \! \le \! l$ are important,
the full quantum problem needs to be solved.

%**************************************************************
\section{Finite temperature effects. Genesis of the QHE}
\label{fintemp}

We now consider how the picture presented in the previous Section
manifests itself in the QHE. Our object of attention will be so-called
clean systems which are experimentally realized in GaAs/AlGaAs
heterojunctions \cite{Efros}. 
In those systems, a random potential is mainly created by remote Silicon
donors placed in a layer parallel to the 2DES, a spacer
thickness $d$ (of the order of several hundred \AA) away from it. 
As demonstrated by Efros \cite{Efros}, 
a random distribution $C({\bf r})$ of such donors produces an
electrostatic potential $\Phi({\bf r})$ in the plane of 2DES which 
is given by: 
\be
        \Phi({\bf r}) = 2 \pi \frac {e^{2}}{\epsilon}
        \int {\frac {d^{2}q}{q} \, C({\bf q})\, e^{i {\bf qr} - qd}} \;,
\label{Random}
\ee
where
\be
        C({\bf q}) = \int {\frac {d^{2} r}{(2\pi)^{2}}
        \, C({\bf r}) \, e^ {-i {\bf qr}}} \;.
\label{Fourier}
\ee
Equation (\ref{Random}) clearly shows that harmonics of the random potential
with $q\:\! d\!>\!1$ are exponentially suppressed. This lead the authors of
Refs.\ \onlinecite{Efros,Efros3,Efros4} to the conclusion that
short-wavelength potential fluctuations are not important. Let us
analyze the picture of the QHE under this assumption.

For filling factors $\nu$ close to an integer, say, from the upper
side, $\nu \!=\! N\! +\!\delta$, there are very few electrons above the
top-most filled Landau level. As shown in Refs.\ \onlinecite{Efros,Efros3},
the transition between the wavelengths of the harmonics that are
effectively screened (linear screening regime) and those which are
poorly screened (non-linear screening regime) can be written as 
\be
        R_{c} = \pi l^{2} \frac {\sqrt{C}}{\delta} \;,
\label{Rcrit}
\ee
where $C$ is an average areal donor density. For sufficiently small
$\delta \,$, $R_{c}$ becomes large and surpasses the spacer
thickness $d$. In this case some strong harmonics in the random potential
are left unscreened since screening would require many more electrons
from the lower Landau levels to
participate and, therefore, many excitations over the magnetic gap
$\hbar \omega_{c}$. If the region with
non-linear screening percolates, the QHE is observed since in the
non-linearly screened regions only completely filled or empty Landau
levels are present. For spin-degenerate Landau levels a rough estimate
gives $\mid \!\delta\! \mid <\! \delta_{c} $, with $\delta_{c}\!\simeq\!
0.1 $ for the width of the QH plateau under typical conditions.
This clearly contradicts with the observed QH plateaus at small
temperatures which are very wide.

Another weakness of the theory, originating in the neglect of 
the short wavelength harmonics of the disorder potential, 
is its prediction for the temperature dependence of the QH plateaus
widths \cite{Efros}. Contrary to the experimentally observed
monotonous increase of the width with lowering the temperature
\cite{Gee} with a feature at a non-universal temperature of about
$T\!\simeq\! 300$mK, this theory predicts narrowing of the plateaus
with cooling down. 

\begin{figure}
  \begin{center}
    \leavevmode
    \epsfxsize=8.5cm%\epsfysize=6cm
    \epsfbox {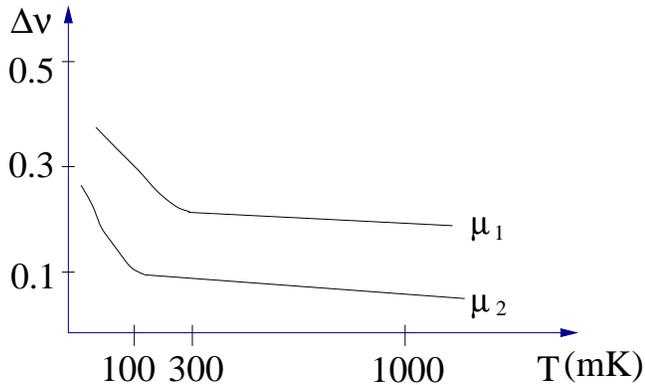}
  \end{center}
  \caption[Dependence of the plateau width on temperature (after Gee
  {\it et al.})] 
  {Temperature dependence of the QH plateau width. Measurements
    for two samples with mobilities $\mu_1 < \mu_2$ are shown (after
    Gee {\it et al.} \cite{Gee}). 
    {\label{Plateaus}}}
\end{figure}

\begin{figure}
  \begin{center}
    \leavevmode
    \epsfxsize=8.5cm%\epsfysize=4cm
    \epsfbox {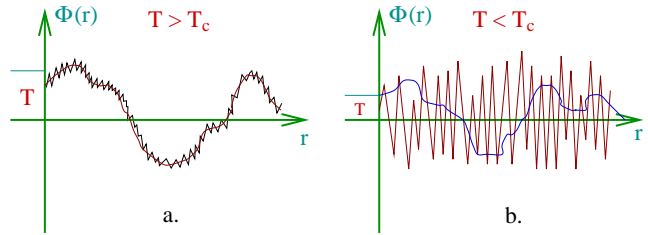}
  \end{center}
  \caption[Long-range disorder potential vs. short-range disorder
  potential above and below the crossover temperature] 
  {Fluctuations of the self-consistent electrostatic potential
    in regions of linear screening. Long wavelength fluctuations
    are screened to a size of the order of the temperature. Short
    wavelength fluctuations are not screened. 
    a.: \,$T > T_{c}$, short wavelength fluctuations are negligible
    compared with $T$;
    b.: \,$T < T_{c}$, temperature is small compared with the
    unscreened fluctuations. 
    {\label{Ripples}}}
\end{figure}

Based on the discussion following Eq.\ (\ref{Qs}) we will show that it
is possible to complement the present theory to build a consistent
description of the QH effect spanning a wide range of temperatures. 
Realistic conditions for observing the
QHE assume some finite temperature $T$. This, of course, affects the
density of states. The sharp peak at Fermi energy is smeared into
a broader line. Consider a filling factor at which the region of
linear screening percolates, $\mid\! \delta \! \mid \ge\!
\delta_{c}$. Then the states 
inside the percolating region correspond to states inside the
DOS peak near the chemical potential $\mu$, and their energies
fall within the temperature $T$ around $\mu$. This description is
correct as long as linear screening is in effect. We have argued that all
harmonics {\it are} linearly screened (in the sense explained above)
as long as the wavelength $\lambda \!\gg\! l$. In other 
words, even inside a linearly screened region, fluctuations of
the disorder potential with sufficiently short wavelengths are not
significantly affected by electron interactions. 
However, if the amplitude of these 
fluctuations is much smaller than the temperature they are
not seen by the system; they appear like small ripples on top of big
waves [Fig.\ \ref{Ripples}(a)] and the occupation numbers of the states
involved are practically unchanged. 

When the size of these unscreened fluctuations becomes
comparable with temperature, however, the distribution of occupation
numbers changes dramatically, and eventually, when the amplitude of
high frequency harmonics exceeds the temperature, the possibility of partial
filling is eliminated. In this case the linearly-screened, metallic
regions, break into pieces of completely filled or empty top-most
Landau level. 
In each and all of these parts, the electrons behave as an
incompressible liquid and depending on which filling factor
percolates, the Hall conductance takes the corresponding quantized
value [$Ne^{2}/h$ or $(N\!-\!1)e^2/h$]. 

It is now possible to make a rough estimate for the crossover
temperature. If we denote the crossover wavelength by $\lambda_{c}$ and 
use Eq.\ (\ref{Random}), we find that the amplitude $V_{c} $ of the
harmonic $\lambda_{c} $ is given by
\be
        V_{c}(q \! = \! 2\pi/ \lambda_{c}) = V_{0} e^{-2 \pi d /
        \lambda_ {c}} \;,
\label{Vc}
\ee
where $V_{0}$ is a typical amplitude of the long-wavelength
fluctuations. Because of the uncertainty in $\lambda_{c} $ and the
exponential dependence of $V_{c}$ on it, the evaluation of $V_{c}$ can
be unreliable. We note, however, that for $\lambda_{c} \!=\! d$, $V_{c} =
2 \times 10^{-3} V_{0}$. Since $V_{0}$ is usually of the order of several
$\hbar \omega_{c}$, and for typical densities $\nu\! =\!1$ corresponds to
$\hbar \omega_{c}\! \simeq \!100 $K, we have an estimate of
several hundred millikelvin for $V_{c}$. As
discussed above, $V_{c}$ has also a meaning of crossover
temperature $T_{c}$, and we therefore conclude that as we lower the
temperature around the value $T \! \sim \! T_{c}$ given
by the right-hand side of Eq.\ (\ref{Vc}), the system
crosses over from dissipative to non-dissipative, quantum Hall
regime. This cross-over temperature is clearly non-universal since it
depends on the magnetic field, the spacer thickness and the strength
of disorder.

We will show below that above the crossover temperature the width of
the QH plateau is almost independent of the temperature with a slow
growth tendency towards lower temperatures. It is clear that upon
crossing $T_{c}$ {\em down} the width of the plateau starts growing
fast since more short-wavelength harmonics become ``seen'' by the system
and larger areas of previously metallic regions break into
incompressible pieces. Such a crossover feature has been seen in
several experiments at temperatures below $1$K
\cite{Komiyama1,Nachtwei,Gee}. At even lower temperatures the
long-range part of disorder potential becomes completely irrelevant
and one can treat the system as composed of independent electrons in a
short-range disorder potential. It is
believed that at zero temperature such a system exhibits sharp
transitions between the QH plateaus \cite{Huckenstein} so that there
is no percolating metallic region at any filling factor.
For sufficiently narrow transitional region between the plateaus the
behavior becomes universal, i.e. independent of the particular disorder
configuration and electron density. This feature is entirely the
consequence of the short-range nature of the random
potential. Therefore, in order to approach the universal regime the
temperature needs to be reduced at least below the crossover
temperature. For very clean samples (for example, with a spacer
thickness of around 1000 \AA \cite{Gee}) $T_{c}$ can become less than
$10$mK and universal behavior is usually not achieved.

At this point, it is important to emphasize that, for the
relatively clean samples under consideration, the QH plateaus are wide
{\em only} at temperatures $T \stackrel{<}{\sim} T_c$, where the
quantum effects mentioned above are dominant. In other words, for a
wide range of magnetic fields, the QH effect is only observed in the
low temperature regime, and as we increase the temperature above
$T_c$, a metallic region percolates, extended states are close to the
Fermi energy and dissipative processes become important. 

Consider now the situation when the filling factor is outside the
plateau at $T > T_c$. The extended state is then in the compressible
region (within  $E_F \pm k_BT$), and all states in the incompressible
region are localized. Let us now consider the situation in the 
low-temperature regime ($T < T_c$). Since states in the formerly
incompressible region are not majorly affected by quantum effects, we
must conclude that any extended state must come from one of the
formerly partially filled states. It immediately follows that the
distance between the energy of extended states and the Fermi level,
or the width of the conventional gaussian DOS in this ``quantum
regime'' cannot exceed $k_B T_c$. 

There is one additional dramatic consequence of the constraints on
the QH liquid in the ``quantum regime'': there is a maximum
``wavelength'' of a few magnetic lengths for an alternating structure
of completely filled and completely empty states. Consider the
transformation of the  {\it compressible} liquid as the temperature is
lowered past $T_c$. Only states in the compressible regions can
vary the occupation numbers, but these variations are constrained by
the necessity to keep Coulomb energy close to the minimum. In order to
achieve this, the occupation numbers must alternate with a period
comparable to the magnetic length, so that the charge density is not
changed significantly. This is in a deep contrast with a conventional
perception of the incompressible QH liquid as homogeneous on almost
macroscopic length scales. 

Similar arguments can be used to study the fractional QHE (FQHE). In
general, one would have to talk in terms of composite fermions
rather than electrons. This would complicate the
picture: the Hamiltonian for composite fermions contains a term
which makes the effective magnetic field dependent on the local
charge density. However, in the incompressible liquid with a quantized
filling factor (i.e. $\nu = 1/3$), the density is constant.
In incompressible regions the disorder potential is unscreened and,
therefore, the electric field in these regions is of the order of the
bare electric field ($\simeq 0.1$V/$\mu$m). Since the
FQHE gap for the strongest fraction does not exceed $1$meV, the width
of the FQHE liquid may not exceed 2--3 $l_B$. This, most probably, makes
it impossible to observe narrow electrostatic-type FQHE plateaus above
the corresponding crossover temperature. It also creates additional
requirements on a consistent theory of the FQHE. Since the FQHE is
thought to be a collective effect, only the liquid with a certain
degree of homogeneity can display the FQHE: the narrow rivers
mentioned above do not satisfy this property. Only few electronic
states across the river may simultaneously be the part of FQHE
liquid. These speculations demonstrate the need for modifications in
the Laughlin state to consistently describe the FQHE in a disordered
system.

Different approach has to be used in ``dirty''
samples realized in InAs/AlInAs heterojunctions or in Si MOSFETS. Due
to the different origin of the disorder potential, the
short-wavelength harmonics are not suppressed 
exponentially and universal behavior can, in general, be observed at
any temperature where the system exhibits the QHE. If a QH plateau
exists in such a system it is always wide, and temperature induced
onset of dissipation occurs at temperatures which have little
dependence on the filling factor within the plateau.

% ....................................................................
\subsection{Melting of the QHE liquid}

 This temperature dependent picture can be extended to the other limit
of temperature $T \gg T_{c}$. Then only long-wavelength fluctuations
of the potential are relevant in both linearly and non-linearly screened
regions. Since the QHE is observed when the latter one percolates, one may
ask a question of what are the possible mechanisms that destroy the
non-dissipative regime as the temperature is increased. The amplitude
of non-linearly screened harmonics is of the order of $ \hbar
\omega_{c}$ while the root mean square of the electric field is the same as
for bare disorder potential ${\cal E}\!=\! 0.1 \hbar \omega_{c} /e l $. If
excitations over the magnetic gap were a relevant mechanism for melting,
the corresponding temperature $T_{m}$ would be of the order of 
$\hbar \omega_{c}$
while the QHE disappears completely at a temperature an order of
magnitude smaller. The following mechanism gives the right order of magnitude
estimate for $T_{m}$. 

\begin{figure}
  \begin{center}
    \leavevmode
    \epsfxsize=8.5cm%\epsfysize=6cm
    \epsfbox {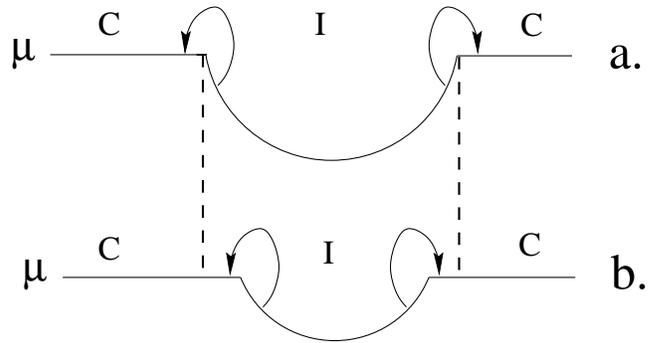}
    \caption [Mechanism of melting of the QH system]
    {Illustration of the melting mechanism of the QH system. $I$
      stands for ``incompressible'', $C$ for ``compressible''. Thermal
      fluctuations increase the energies of states near the
      compressible regions. 
      {\label{Melting}}}
  \end{center}
\end{figure}

Consider two isolated metallic lakes separated by
an incompressible region (Fig. \ref{Melting}). 
Energies of the states in the lakes
can be considered to be equal to the chemical potential $\mu$;
states inside the incompressible region are completely occupied and
have energies lower than $\mu$. Consider a state in the
incompressible region nearest to one of the lakes
[Fig. \ref{Melting}(a)]. Its
energy relative to $\mu$ can be estimated
as $\epsilon \!=\! -e {\cal E} l$, since the average extent of a state is
of the order of the magnetic length $l$. 
If the temperature $T$ is comparable to $\epsilon$, the state
becomes partially occupied and thermal fluctuations force its energy
to raise closer to the chemical potential. This can happen only due to
interactions with other electrons both in the lake and in the
incompressible region. As a result of these interactions the size of
the completely filled region shrinks and all states raise their
energies as well. This situation repeats now with the new
nearest-to-the-lake state [Fig. \ref{Melting}(b)]. 

It might happen that, during such a chain process, the electric field
close to the boundary increases. Then the process is locked in this
region unless the temperature is further increased. A natural estimate
for the thermal breakdown temperature of the QHE is then 
\be
        T_{m} = e\,{\cal E} \, l = 0.1 \, \hbar \omega_{c} \;.
\label{Td}
\ee
This formula gives a range from a few to several degrees Kelvin, 
% between $2$ and $20$K for $T_{m} $ 
in clean samples, in qualitative agreement with 
most experiments \cite{klitz1,Komiyama1}. 

% .................................................................
\subsection{Plateau width dependence on the number of the filled
Landau levels}

It is intuitively clear that the width of the QH plateau should be a
decreasing function of the number of the filled Landau levels. To
understand such dependence we notice that it is the charge density
(not the filling factors) which is responsible for screening of the
external potential. Consider the same sample at different magnetic
fields: in the first case the completely filled first LL percolates,
in the second it is the filled second LL ($\hbar \omega_c^{(1)} > \hbar
\omega_c^{(2)})$. The charge density distribution in the first case leads
to the unscreened fluctuations of the potential of the order of $\hbar
\omega_c^{(1)}$. However, such deep fluctuations are impossible in the
second case: as soon as the potential falls below $\hbar
\omega_c^{(2)}$, the 3rd LL starts filling. Therefore, the regions
with partially filled 3rd LL at smaller magnetic fields are larger in
size 
than the corresponding compressible regions with partially filled 2nd
LL at stronger magnetic fields. Additional electrons on the top 3rd LL
must have come from some other regions in the system: overall it
should be neutral on sufficiently large length scales. The only
possible 
reservoir for such electrons are the percolating incompressible
regions. By reducing the density, these incompressible regions
decrease in size and part of them become metallic. We conclude that
both the regions with partially filled 2nd and 3rd LL's are larger
than their counterparts (1st and 2nd LL's respectively) at stronger
magnetic fields. Therefore, the QH plateau becomes narrower with
decreasing the magnetic field, as expected.

% ...............................................................
\subsection{Density of states as a function of temperature}

We now return to the DOS and seek to understand how it changes when
the temperature is varied. At high
temperatures, $ T \!\gg\! T_{c}$, there is a peak around the
Fermi energy with a width of the order of temperature
(Fig.\ \ref{DOS}). 
The amplitude of any unscreened harmonic in the linear screened
regions is negligibly small, as compared to the temperature and cannot
be seen inside the resonance curve. The tail in the DOS corresponds to
states in the incompressible regions.  

For temperatures $T_{c}\! \ll\! T \ll T_{m}$ this DOS is
practically temperature independent (except for a rounding of the peak
at higher temperatures). According to the previous discussion,
as $T$ approaches and crosses below $T_{m}$, more states from the tail
of the DOS are pumped into the peak. The peak width,
of the order of the temperature, changes very little compared to the
total width of the DOS. Therefore, in this range of temperatures, the
DOS changes by increasing the height of the peak at the expense of the
reduction of the weight of the tail. 

On the other side, for $T\! \ll \! T_{c}$, the amplitude of
unscreened fluctuations becomes greater than the 
width the peak would have at this temperature in the absence of
short-range disorder. The width of the peak in the DOS in this case is
determined by the strength of short-wavelength disorder potential and is
independent of temperature. This is a common non-interacting symmetric
DOS with a few modifications: the existence of a residual DOS at the
tails spreading out to energies comparable with 
$\hbar \omega_{c}$, and the distance between the line of symmetry of
the DOS and the Fermi level is of the order of 
$T_c \ll \hbar \omega_c$. The Fermi energy is either below or  
above the center of the peak depending on the filling factor.   

\begin{figure}
  \begin{center}
    \leavevmode
    \epsfbox {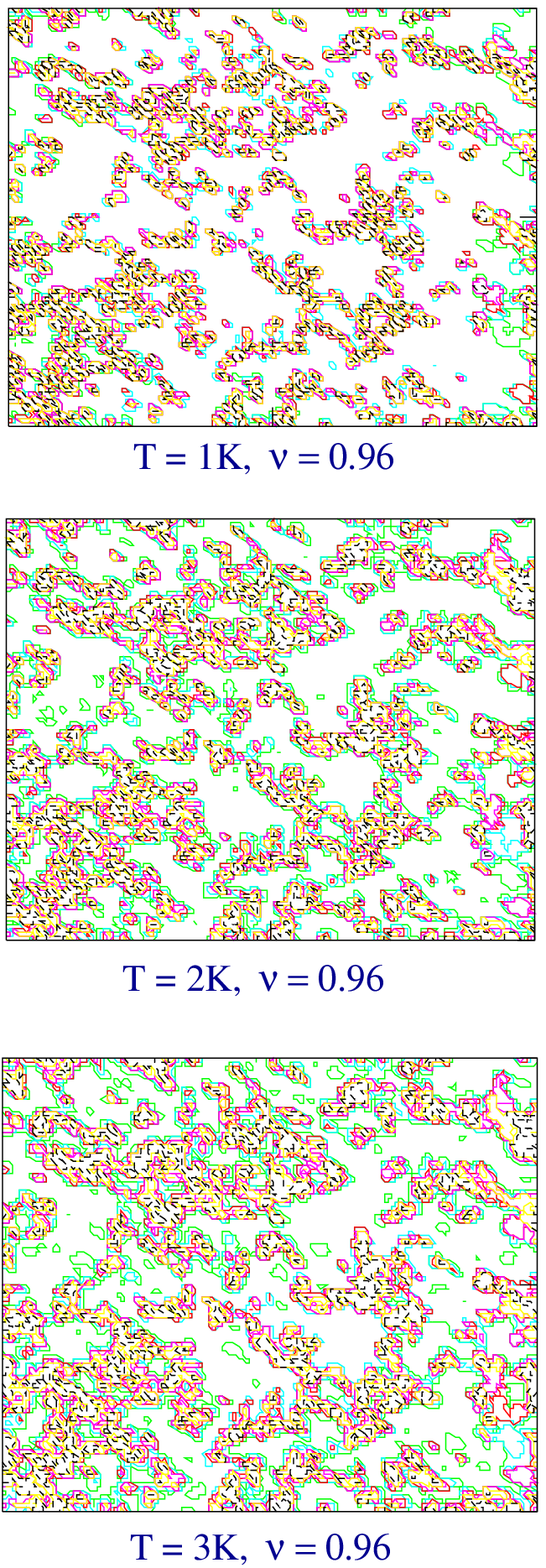}
  \end{center}
  \caption[Melting of the QH system. Numerical simulations.] 
  {Numerical simulations of the melting of the QH system. The same
    sample, with a filling factor $\nu=1.92$, is studied at different 
    temperatures. Shaded areas are compressible, white areas are
    incompressible. a. $T=1$K; b. $T=2$K; c. $T=3$K. As the
    temperature is raised compressible areas finally percolate,
    destroying the QH regime.
    {\label{Melt_num}}}
\end{figure}

We were able to simulate numerically the state of the system  for 
$T\!>\! T_{c}$.  The procedure consists of minimization of
the free energy with respect to local density at finite
temperature. The free energy of the system can be written as 
\ba
        {\cal F} = {\cal H}_{1} + && k_{B} T \int  d^2 r
        \sum_{N} \left(  n^{(N)} ({\bf r}) \log [
        n^{(N)} ({\bf r})] \right.  + \nonumber \\
        && \left. + [ 1- n^{(N)} ({\bf r})] \log [ 1 - n^{(N)}
        ({\bf r})]\right)  \;,
\label{Freen}
\ea 
where ${\cal H}_{1}$ is defined in Eq.\ (\ref{Elst}). The last
term in Eq.\ (\ref{Freen}) is the entropy of the system described by Fermi
statistics. Fig. \ref{Melt_num} demonstrates the density distribution in
clean system with average occupation number around $\nu\!
=\!1.92$. Part (a) corresponds to a temperature $T \!=\! 1K \!< \!T_{m}$.
Figure \ref{Melt_num}(b) corresponds to a slightly
higher temperature, $T=2$K: some expansion of the compressible region
is observed. Finally, Fig.\ \ref{Melt_num}(c) represents
the same system at $T=3$K which is apparently greater than $T_{m}$. We
could bracket $T_{m}$ to within an interval between $2.5$K and
$2.8$K. We also observed very weak dependence of $T_{m}$ on the
filling factor. An interesting result produced by these calculations is 
that when the melting occurs at $\nu \!\ge\! 2.0$,
a metallic percolating region is formed both on the first and second
spin-degenerate Landau levels practically at the same time. All the
features discussed above are well seen on
these plots. Unfortunately, the electrostatic approximation by its nature
does not allow us to probe the system at temperatures around and below
$T_{c}$. A more sophisticated approach is needed to quantify and,
probably, numerically illustrate the crossover picture. At the moment
we are unable to present a satisfactory, quantitative model for the
microscopical nature of the crossover regime as well as microscopic
picture of inhomogeneous QH liquid at temperatures below $T_{c}$ (see
Refs.\ \onlinecite{Pryadko1,ShklFogl,Shimshoni}).

% ===========================================================
\section{Dissipative processes and energy relaxation in the QH regime}
\label{heat}

The characteristics of the DOS at temperatures above $T_{c}$ has various
implications on the dissipative dynamics in QH systems. We will
concentrate on one of them, namely, the energy relaxation rate in the
QH regime. By that we imply that we consider a ``clean'' system at
$T \!>\! T_{c}$ and at such filling factors that the Hall conductance is
quantized with high accuracy. In this Section we intentionally avoid
calling this regime {\it non-dissipative} since it is dissipation,
however small it might be, that we are interested in. 

Assume that current is injected into the system. Since dissipative
conductivity is finite, there is finite Joule heat released into the
electronic system. Depending on the rate of relaxation of this energy
from the electrons to the lattice, the 2DES will either be in
quasi-equilibrium ($T_{e}\!\simeq\! T_{l}$), or the electron
temperature will raise ($T_{e}\!>\! T_{l}$) to enhance this relaxation
and achieve a steady state. 

The problem of overheating of the 2DES in such systems has been addressed
before in the study of different phenomena. Chow {\it et al.}
\cite{Girvin1} considered this effect in the {\it transitional} regime
between QH plateaus, and argued that the effective electronic temperature
dependence on the current contains needed information to reproduce one
of the scaling exponents in the universal regime. Although the
temperatures studied in Ref.\ \onlinecite{Girvin1} are apparently low
(between 100mK and 500mK) they may still be higher than $T_{c}$ since
the sample they considered had a thick spacer ($d=1800$\AA). The 
authors argued that the energy relaxation rate to acoustic
(piezoelectric) phonon modes scales, at these low temperatures and
strong magnetic fields, as $T_{e}^{4}$; it was implicitly assumed that
the energy relaxation within the 2DES due to inter-electron interaction was
sufficiently fast and (with some justification) that diagonal
conductivity $\sigma_{xx}$ was independent of wavevector $q$ and
electron temperature $T_{e}$. The latter assumption obviously fails in
the QH regime since finite conductivity results from either thermal
activation ($\sim \!\exp[{\epsilon_{a}/k_{B}T_{e}}]$) or variable range
hopping ($\sim \! \exp[{-A/T_{e}^{\beta}}]$), where $\epsilon_{a}$ is an
activation energy, and $\beta$ is some model-dependent positive power. 

Our analysis of heating at filling factors within the QH
plateaus is reminiscent of the analysis of Ref.\
\onlinecite{Girvin1}. However, we will use explicitly the 
properties of the DOS, since it is unclear how to express the surface
acoustic wave absorption probability through a global conductivity in
the case of spatially inhomogeneous system (see also Ref.\
\onlinecite{Galperin}). 

The notion of overheating of the electron system assumes that the
electrons are considered to be in thermal equilibrium at some temperature
$T_{e}$. This implies that all the energy pumped into the system is
distributed throughout it at rate $\tau_{e}^{-1}$ which is much
greater than the rate of any other energy relaxation process in the
system. Important processes which are 
to be considered are the supply of energy from some external source and
release of this energy to a thermal bath. If $\tau_{e}$ is the shortest
time scale in the problem, and if the temperature of the
thermal bath is denoted as $T_{b}$, the 2DES temperature $T_e$ is
defined by the energy balance equation
\be
        P_{in} = P_{\rm out}(T_{e}) \;,
\label{Te}
\ee
where $P_{in}$ is the input power, and $P_{\rm out}$
is the energy relaxation rate from the electrons to the heat bath. The
relevant question is how much the electron temperature needs to be
raised so that Eq.\ \ref{Te} is satisfied. If energy relaxation to the
lattice is slow, it is clear that considerable overheating ($T_e
\!\gg\! T_b$) will be observed. Conversely, a fast relaxation
mechanism assures that both temperatures are similar ($T_e
\!\simeq\! T_b$).

The energy fed into the 2DES QH system is Joule heat associated with
the injected current $j$ and dissipative resistivity $\rho_{xx}$:
$P_{in} \!=\! \rho_{xx}j^{2}$. This assumes that most dissipative
processes do not involve electron-phonon interactions, otherwise part
of the heating is directly applied to the lattice. At this point it is
important to emphasize that in the QH regime $\rho_{xx}$ is 
exponentially small. For the phonon emission power we write:
\ba
        P_{\rm out}(T_{e}) = 2 \pi \sum_{i,f} \sum_{\bf
        q} \hbar\omega({\bf q}) 
        \mid \! M_{i,f}({\bf q}) \! \mid ^{2}  f_{i}\, (1\! - \!
        f_{f}) \times \nonumber \\
        \left[ \,(n_{\bf q} \! + \! 1) 
        \,  \delta[\epsilon_{i} \! - \! \epsilon_{f}\! -
        \! \hbar \omega ({\bf q})] - 
        n_{\bf q} \, \delta[\epsilon_{i} \! - \!
        \epsilon_{f} 
        \! + \! \hbar \omega ({\bf q})] \right] \;.
%
%       P_{\rm out}(T_{e}) = 2 \pi \sum_{i,f} {\sum_{\bf
%       q}{\hbar\omega({\bf q}) 
%       \mid D_{i,f}({\bf q}) \mid ^{2} \left (f_{i}(T_{e}) (1 -
%       f_{f}(T_{e})) n_{\bf q} \delta(\epsilon_{i} - \epsilon_{f} -
%       \hbar \omega ({\bf q})) -
%       (1 - f_{i}(T_{e})) f_{f}(T_{e})(n_{\bf q} + 1) 
%        \delta(\epsilon_{i} - \epsilon_{f} +
%       \hbar \omega ({\bf q})) \right)
%       }}
\label{Emission}
\ea 
where indices $i$ and $f$ correspond to initial and final electronic
states respectively, ${\bf q}$ is the phonon momentum; $f_{i,f}(T_{e})$ are
fermionic filling factors at temperature $T_{e}$, $n_{\bf q}(T_{b})$ is
a bosonic filling factor at lattice temperature $T_{b}$, and $M_{i,f}({\bf
q})$ denotes the matrix element of the electron-phonon interaction. In
Eq.\ (\ref{Emission}) the first term inside the vertical brackets
stands for phonon emission while the second corresponds to phonon
absorption. To estimate $P_{\rm out}$ we make the following
approximation: each electronic state can be visualized as a strip with
a transverse size of the order of the magnetic length $l$ stretched along
some equipotential. Inside any individual potential fluctuation this
potential can be linearized. The wave function in this linearized
region can then be written as a solution of Landau equation in a
homogeneous electric field. Solutions of this equation are enumerated
by the local set of one-dimensional wave vectors $k$ that have meaning
of the position of the wave functions center \cite{Feng}, which for
the lowest Landau level read
\be
        \psi_{k}(x,y) = \frac{1}{\pi^{1/4} \, l^{1/2}} exp \left[
        ikx - \frac{(y-y_{0})^2}{2 l^{2}} \right] \;,
\label{Wavef}
\ee
where $x$ is the direction along the trajectory and $y_{0} \simeq
kl^{2}$ is the guiding center position. Energies of these
states follow the external potential in the fluctuation:
\be
        \epsilon_{k} = \frac{1}{2} \hbar\omega_{c} + k l^2 e {\cal E}\;,
\label{Energy}
\ee
with $\cal E$ being the typical electric field inside the
fluctuation. Since the 
characteristic wavelength of potential fluctuations in the regime we
are studying is much larger than $l$, and any transition due to
electron-phonon interaction is possible as long as initial and final
states have substantial overlap, such transition occurs when the electron
moves from one strip to another approximately $l$ away in the same
potential fluctuation. 
Using the dispersion relation for longitudinal acoustic
phonons $\omega \!=\! c_{L} q$, with 
$q \!=\! [q_{x}^2 \! + \! q_{y}^2 + \! q_{z}^2]^{1/2}$, an estimate of
${\cal E} \!\simeq\! 0.1 \, \hbar \omega_{c}/e l$, and given that  
$c_{L} \!\sim\! 10^{3}$ m/s, one can make sure that single phonon
processes are indeed allowed: 
$e {\cal E} \triangle y_{0} \!=\! c_{L}q \!>\! c_{L}q_{x} \!\sim\!
 c_{L} \triangle y_{0} /l^{2}$ leads to ${\cal E} \!>\! 0.005 \, \hbar
\omega_{c} /e l$, which is certainly the case. The matrix element
$M_{i,f}({\bf q})$ calculated with the wave functions from Eq.\
(\ref{Wavef}) depends on the wave vector of initial state only through
a phase factor that is irrelevant in this problem.  

Since each strip passes through many potential fluctuations
(provided that the corresponding equipotential is sufficiently long) the
electron on the strip probes many random configurations of the
potential and, therefore, averages over the whole ensemble. Hence, one
can use the ensemble averaged density of states $D(\epsilon)$ to
describe the probability of the electron having particular initial and
final energies, at least for electrons in the incompressible
regions. In this approximation, Eq.\ (\ref{Emission}) takes the form:
\bea
\label{Nice}
        P_{\rm out} && = 
        2 \pi \sum_{{\bf q}} \mid \! M({\bf q})\! \mid^{2}
        \frac{1}{\alpha - 1} \times
          \\
&&  \!\!\!\!\!\! \sum_{\epsilon_{\bf k}}    \left[ \alpha x \, 
          \frac{ D(\epsilon_{k})D(\epsilon_{k}-c_{L}q)}
          { (\beta + x)(1 + x) } - \beta  x \, 
          \frac {D(\epsilon_{k})D(\epsilon_{k}+c_{L}q)}
          {(\beta x + 1)(1 + x)} \right]  \;, \nn
%
%       {1 + e^{\epsilon_{k}/kT_{e}}}
%       \frac {e^{c_{L}q / kT_{b}}{e^{c_{L}q / kT_{b}}-1}
\eea
where we denoted $x \!=\! \exp[{\epsilon_{k}/kT_{e}}]$, 
$\alpha \!=\! \exp[{c_{L}q / kT_{b}}]$ and 
$\beta \!=\! \exp[{c_{L}q / kT_{e}}]$. Changing to an integral 
over $\epsilon_{k}$ and after simple manipulations we obtain:
\bea
        P_{\rm out} =&&  k_{B} T_{e} \sum_{\bf q} 
        \mid \! \tilde M({\bf q}) \! \mid ^{2} 
        \frac{\alpha - \beta}{\alpha - 1} \times \nn \\
        && \times  \int_{0}^{\infty} \!\! 
        dx \, \frac{ D(k_{B}T_{e}\log{x}) 
          D(k_{B} T_{e} \log {x / \beta})}
        {(\beta + x) (1 + x)}\;,
\label{Finalle}
\eea
where we absorbed all numerical factors into a redefinition of the matrix
element $\tilde M({\bf q})$. We did not have to specify the exact form of
electron-phonon interaction; it can, in fact, be of either piezoelectric
or deformation potential nature. The relative contribution of these two
types of interaction depends on the temperature \cite{Price}, however
both are of the same order of magnitude at temperatures around
$1$K. After averaging over phonon polarization, one gets matrix elements
of the order of unity. We chose not to go into details of the form of
these matrix elements (see Refs.\
\onlinecite{Price,Toombs,Dietzel,Tamura}) since for our purposes only
the order of magnitude matters. We kept the factor $k_{B}T_{e}$ as it
appeared with the change of integration variable for the sake of
dimensionality.
 
The most important part defining the magnitude of $P_{\rm out}$ in
Eq.\ (\ref{Finalle}) is the density of states $D(\epsilon)$. For example,
if only states well below the Fermi level, with energies $\epsilon
\ll k_{B}T_{e}$, have weight in the DOS then the range of integration
over $x$ in Eq.\ (\ref{Finalle}) becomes exponentially narrow: the
upper limit of integration is then of the order of
$\exp[{-\epsilon_{0}/k_{B}T_{e}}]$, where $\epsilon_{0}$ is the highest
energy such that $D(\epsilon_{0})$ is not exceedingly small. 
On the other hand, if the DOS 
spreads all the way to the Fermi level, the corresponding integral in
Eq.\ (\ref{Finalle}) is not small nor is the emitted power $P_{\rm
out}$. 

To see that energy relaxation rate due to phonon emission is
sufficient to keep the difference $T_{e} - T_{b}$ negligibly small, we
recall that Joule heating rate is suppressed exponentially by a factor of
$\exp[-\epsilon_{a}/kT_{e}]$: dissipative transport is due to the
electrons (or holes) in the extended states. Different mechanisms of
scattering are available for these electrons. Scattering off the
electrons in metallic, compressible regions is one mechanism leading
to thermal equilibration within the 2DES. The other mechanism proposed
in \cite{Shklov1,Shklov2} involves interaction with lattice phonons. It
is not clear which of these contributions to dissipative resistivity
prevails. If under certain conditions the latter one dominates, the
2DES has to be in thermal equilibrium with the lattice and
$T_{e}\!\simeq\!T_{b}$. Electron-electron scattering is the only way
of releasing energy into 2DES. It is clear, however, from Eq.\
(\ref{Finalle}), as well as from intuitive arguments, that 
electrons in metallic  regions are interacting with the phonons very
effectively and are never a bottleneck for thermal equilibration with
the lattice. In general, the metallic regions are always in thermal 
quasi-equilibrium with the lattice. 
They may not be equilibrated with the electrons in
incompressible regions: in this case the QH system has to sustain the
temperature gradients. However, it assumes that the dissipation of
energy by incompressible electrons carrying the Hall current occurs
via interaction with phonons: in the QH system diffusion (and
therefore dissipative conductivity) occurs through inelastic
scattering due to the non-degenerate nature of the states in
long-range disorder potential.

Expression (\ref{Finalle}) is applicable in the transitional regime as
well. However, when the metallic region percolates, dissipation is also
greatly enhanced. Both Joule heating and phonon emission are
determined in this case by the same DOS, that is by a finite fraction
of the total number of states at the Fermi level. Observation of
overheating in this regime \cite{Girvin1} is therefore not
surprising. 

%{\bf Here I'll add some discussion of diffusion in these systems and
%of the possibility of scattering by the short-range disorder
%potential. Some comments on the mechanisms of quantum diffusion vs. 
%mechanisms of conductivity.}  

% ===========================================================
\section{Conclusions}

In this paper we demonstrate that screening of long-range fluctuations of
the disorder potential leads to a novel form of the density of states.
Contrary to the conventional DOS for non-interacting electrons, this
screened DOS possesses a distinctive peak at the Fermi level
originating from states in metallic (compressible) regions. We argue
that this form of DOS remains almost intact with changes in the magnetic
field, the only variation being in the width of the peak and in the
position of the energy of the extended states relative to the Fermi
level. We also argue that, as the temperature of the system crosses
below a certain value $T_c$, short-range fluctuations of the disorder
potential become dominant and the nature of the states within the
narrow peak in DOS changes. This $T_c$ is the lower boundary of the
range of validity of the electrostatic approximation. 
 
For filling factors that only exhibit the QHE at low temperatures 
($T \stackrel{<}{\sim} T_c$), short-range fluctuations dominate and
the full quantum problem needs to be solved.
Based on numerical and analytical results obtained by a
number of authors, one can expect that for non-interacting electrons,
the DOS has symmetric gaussian shape, centered at the energy of extended
states. In light of our developments, the assumption of
non-interacting electrons needs to be amended by the requirement that
the charge density remain unchanged when the temperature is lowered
through $T_c$, that is, it is still prescribed by
electrostatic solution even at low temperatures. This leads not only to a
constraint on the distance between the energy of extended states
and the Fermi level never to exceed $T_c$ but also to a more dramatic
conclusion: namely, that in these, short-range fluctuations dominated
regime, the percolating quantum Hall liquid can nowhere be wider than
few magnetic lengths. This conclusion calls for a closer look at the
existing quantum theory of the QHE which is based on the assumption of
an homogeneous system. 

The current common perception of the QHE is that it is 
an entirely quantum phenomenon. In this sense, the theory of the QHE
refers to the lower temperatures ($T<T_c$) and to the magnetic fields
away from the center of the plateaus. Yet it is important to bear in
mind that this QH system is very inhomogeneous and that the ``pure''  
considerations do not apply to the whole system but rather to the
regions where the disorder potential is metallically screened. All the 
observed quantum effects, including the universal behavior at low
temperatures, have their origin in these regions. The complete theory
of the IQHE has to account for this non-trivial microscopic structure 
of the 2DES.

Based on the understanding of the QH system described by the
electrostatic approximation, we also present a model of melting of
QH liquid at higher temperatures.  We were able to explain why the
electrostatic solution predicts narrow quantum Hall plateaus while
experimentally observed plateaus are usually much wider, this
difference originating in the crossover behavior described above.

Finally, we applied the derived form of the DOS to the study of
dissipative processes in QH system. We showed that since Joule heating
is limited in the QH effect by a very low dissipative conductivity,
and the relaxation of energy due to the coupling to phonons is very
effective (because of the large number of states on the Fermi
level), electronic overheating cannot occur (or is negligible) in the
QH regime.

% ===========================================================

\acknowledgements

The authors are grateful to Prof.\ David Thouless for his attention to
this work. We thank Jung Hoon Han for numerous helpful discussions
throughout this work, and A. L. Efros for useful discussions at the
early stages of development. This work was supported by the NSF, Grant
No. DMR-9528345. CW was supported in part by Grant No. DMR-9628926. 

% ===================================================================

% ===========================================================
% ===========================================================
%\end{multicols}
% ===========================================================
\end{document}